%% file: main.tex
\documentclass[11pt]{article}
\usepackage[T1]{fontenc}

\usepackage[margin=1.1in]{geometry}   % Reasonable margins
\usepackage{lineno}                 % Line numbering
\usepackage{setspace}              % Line spacing
\usepackage{graphicx}             % For including graphics
\usepackage{hyperref}             % Hyperlinks in the PDF
\usepackage{titlesec}             % Optional for section formatting
\usepackage{authblk}               % For author/affiliation formatting
\usepackage{textgreek}

\usepackage{caption}
\usepackage{placeins}       % FloatBarrier

\usepackage{amssymb}
\usepackage{amsmath}
\usepackage{amsfonts}

\usepackage{siunitx}
\usepackage{physics}
\AtBeginDocument{\RenewCommandCopy\qty\SI}

\usepackage[
  backend=biber,        % Use biber for full BibLaTeX features
  style=numeric,        % Numbered citations
  sorting=none,         % Sort citations in order of appearance
  maxbibnames=2,        % Show up to 3 names in bibliography
  minbibnames=2,        % Force et al. after 3 names
  maxcitenames=1,       % Show 1 name in in-text citation
  mincitenames=1,       % Force et al. after 1 name in citation
  giveninits=true,      % Use initials for given names
  doi=false,            % Hide DOI
  url=false,            % Hide URLs
  isbn=false,           % Hide ISBNs
  eprint=true,          % Show arXiv eprint info
]{biblatex}
\input{bibmacros}

\title{A Spin-Photon Interface in the Telecom C-Band with Long Hole Spin Dephasing Time}

\author[1]{Johannes~M.~Michl\thanks{
    Corresponding authors: \texttt{johannes.michl@uni-wuerzburg.de; reza.hekmati@uni-wuerzburg.de}
    
    ~~These authors have contributed equally to this work.}}
\author[1]{Reza~Hekmati$^*$}
\author[1]{Mohamed~Helal}
\author[1]{Giora~Peniakov}
\author[1]{Yorick~Reum}
\author[1]{Jochen~Kaupp}
\author[1]{Quirin~Buchinger}
\author[1,2]{Jaewon~Kim}
\author[1]{Andreas~T.~Pfenning}
\author[2]{Yong-Hoon~Cho}
\author[1]{Sven~Höfling}
\author[1,3]{Tobias~Huber-Loyola}

\affil[1]{Julius-Maximilians-Universität Würzburg, Physikalisches Institut, Lehrstuhl für Technische Physik, Am~Hubland, 97074~Würzburg, Deutschland}
\affil[2]{Department of Physics, Korea Advanced Institute of Science and Technology (KAIST),
34141 Daejeon, Republic of Korea.}
\affil[3]{Karlsruhe Institute of Technology, Institute of Photonics and Quantum Electronics (IPQ) and Center for Integrated Quantum Science and Technology (IQST).}

\date{}

\begin{document}

\maketitle

% Abstract
\input{Sections/0_Abstract}  % \input instead of \include prevents page break 

% Add sections
\include{Sections/1_Introduction}
\include{Sections/2_Results}
\include{Sections/3_Discussion}
\include{Sections/5_Methods}
\input{Sections/6.1_Acknowledgements}
\input{Sections/6.2_Declarations}

%% Bibliography 
\printbibliography

% Supplementary Material
\input{Sections/Supplement}

%%% END
\end{document}

%% file: bibmacros.tex
\DeclareFieldFormat{title}{#1}
\DeclareFieldFormat[article]{title}{#1}  % no italics, no quotes
\DeclareFieldFormat[book]{title}{\textit{#1}} % keep italics for books
\DeclareFieldFormat[@inproceedings]{title}{\textit{#1}} % keep italics for books
\DeclareFieldFormat[article]{volume}{\textbf{#1}}

\DeclareFieldFormat[article]{number}{\mkbibparens{\textbf{#1}}}

\renewbibmacro*{volume+number+eid}{%
  \printfield{volume}%
  \iffieldundef{number}
    {}
    {\addnbspace\printfield{number}}%
  \setunit{\addcomma\space}%
  \printfield{eid}}

\DefineBibliographyStrings{english}{
  andothers = {\mkbibemph{et\addabbrvspace al\adddot}}
}

\renewbibmacro{in:}{}

%% file: Sections/0_Abstract.tex
\begin{abstract}
Matter qubits that maintain coherence over extended timescales are essential for many pursued applications in quantum communication and quantum computing. Significant progress has already been made on extending coherence times of spins in semiconductor quantum dots while interfacing them with photons in the near-infrared wavelength range. However, similar results for quantum dots emitting at the telecom range, crucial for many applications, have so far lagged behind. Here, we report on InAs/InAlGaAs quantum dots integrated in a deterministically placed circular Bragg grating emitting at \SI{1.55}{\micro \meter}. We quantify the g-factors of electrons and holes from polarization-resolved measurements of a positive trion in an in-plane magnetic field and study the dynamics of the ground-state hole spin qubit. We then herald the hole spin in a pulsed two-photon correlation measurement and determine its inhomogeneous dephasing time to $T_{2}^{*}=(15.9 \pm 1.7)$ \SI{}{ns}. 
\end{abstract}

%% file: Sections/1_Introduction.tex
\section{Introduction}

The phenomenon of quantum superposition is nowadays most prominently featured in the concept of qubits – the basic building blocks of quantum information applications like quantum computing and quantum communication.  While the term \textit{flying} qubit almost exclusively refers to photons~\cite{Slussarenko.2019}, \textit{stationary} qubits can be realized in systems such as trapped ions \cite{Monroe.2021, Hilder.2025}, superconducting circuits \cite{Krantz.2019, Kjaergaard.2020}, nitrogen-vacancy (NV) centers in diamond \cite{Jelezko.2006} or semiconductor quantum dots (QDs) \cite{Loss.1998}.

An interface between flying and stationary qubits allows for local operations on the stationary qubits and quantum information transfer between them using the flying qubits. In solid-state, interfacing stationary qubits with photons is possible due to spin-dependent optical transitions – as can for instance be found in NV centers or QDs – that connect the polarization of an emitted photon with the state of a confined spin \cite{Gao.2015}. This connection leads to non-classical correlations which most notably manifest in the entanglement between a solid-state spin and a photon \cite{Togan.2010, Gao.2012, deGreve.2012}. Based thereon, a work by Lindner and Rudolph from 2009 \cite{Lindner.2009} proposed to utilize the spin-qubit as an entangler, enabling the generation of a linear cluster state of entangled photons. These one-dimensional entangled states are highly sought-after, since they can be fused to high-order \textit{graph states} \cite{Thomas.2024} which, in turn, are the resource states for measurement-based quantum computing \cite{Raussendorf.2003} as well as for all-photonic quantum repeaters \cite{Azuma.2015, Buterakos.2017}. 

%%% Cluster State Platform Comparison and Spin Coherence in NIR QDs
A linear cluster state from a quantum dot has first been demonstrated in 2016 \cite{Schwartz.2016} in the near-infrared (NIR) range below \SI{1}{\micro \meter} and has since then been repeatedly reported for NIR quantum dots \cite{Istrati.2020, Li.2020}, also featuring indistinguishable photons \cite{Cogan.2023, Huet.2025}. Despite this progress, the largest cluster states achieved to date come from neutral atoms \cite{Thomas.2022, Thomas.2024} in cavities, as their isolation from the environment allows for coherence times of \(>\)\SI{1}{\milli \second}. In solid-state emitters, by contrast, the interaction with the environment is limiting the coherence times, although through nuclear spin cooling \cite{Jackson.2022, Nguyen.2023} and advanced decoupling protocols \cite{Zaporski.2023, Appel.2025} values \(>\)\SI{100}{\micro \second} are attainable. Yet these coherence-enhancing methods have not been combined with cluster-state generation, leaving quantum-dot implementations effectively constrained by the inhomogeneous dephasing time \(T_{2}^{*}\) of the spin qubit. There, state-of-the-art values are obtained in the NIR-range, reaching up to \(T_{2}^{*}>\)~\SI{70}{\nano \second} for a hole spin \cite{Huthmacher.2018}. 
 
%%% Introduction of Telecom as Favoured WL
At the same time, practical considerations for long-distance quantum communication and integration with silicon photonics motivate a shift toward emission in the telecom C-band at \SIrange{1.53}{1.56}{\micro \meter}, where optical-fiber losses are minimal. Although NIR-range photons can be frequency-shifted via down-conversion \cite{deGreve.2012, Pelc.2012, Zaske.2012}, the process is intrinsically probabilistic—conversion succeeds only with finite probability, typically introducing additional timing jitter—and is furthermore limited in efficiency, making it difficult to reach the thresholds required for scalable architectures and quantum advantage. Thus, while NIR quantum dots currently offer the most favorable spin-coherence properties, telecom-band operation remains the regime of highest relevance for scalable quantum-network applications.

%%% Indistinguishable Single Telecom Photons from CBGs
Substantial improvements in the optical quality of telecom-emitting quantum dots have been achieved through refined growth techniques that optimize material properties. In recent works we have reported on the enhancing effect of e.g. ternary digital alloying and the optimization of the Ostwald-ripening time on photon properties \cite{Kaupp.2023, Kim.2025, Hauser.2025}. Furthermore, much progress has been made on the integration of telecom QDs into circular Bragg-gratings (CBGs), leading to Purcell-enhanced single photon emission with high brightness and short radiative lifetimes \cite{Kaupp.2023}, as well as indistinguishability \cite{Joos.2024, Holewa.2024, Kim.2025, Hauser.2025}. 
%%% Spin Control and Entanglement at Telecom
Full coherent control of a QD spin qubit emitting telecom photons has been achieved only a few years back \cite{Dusanowski.2022} followed by spin-photon entanglement measurements \cite{Laccotripes.2024}, both featuring modest spin coherence times \(<\)\SI{1}{\nano \second}. Only recently, slightly higher values \(<\)\SI{5}{\nano \second} could be reported utilizing quasi-resonant excitation schemes \cite{Laccotripes.2025, Peniakov.2025, Wasiluk.2025}. 
Until now, studies of telecom-emitting quantum dots hosting spins with inhomogeneous dephasing times long enough to support large photonic cluster-state generation have been absent.

%%% Our work
Here, we report on measurements of a hole spin from an InAs/InAlGaAs quantum dot integrated into a circular Bragg grating with a \(T_{2}^{*}\) exceeding \SI{15}{\nano \second}. The quantum dots studied here, in combination with a tailored CBG design for enhanced emission around \SI{1.5}{\micro \meter}, have shown excellent optical qualities including record values of two-photon interference visibility of \(>90 \%\) \cite{Kim.2025, Hauser.2025}. We utilize p-shell quasi-resonant excitation and exploit the polarization memory of a positive trion (\(X^{+}\)) to study the excited-state spin evolution as well as the coherence dynamics of a ground-state hole-qubit.

%% file: Sections/2_Results.tex
\section{Results}
\label{sec:results}

The data we report on here focus on the optical transition of a positively charged excitonic complex $X^{+}$, also referred to as a positive trion. Basic characterization and trion identification were conducted performing power- and polarization series (cf. supplementary section \ref{sec:app_power_pol_series}). We measured InAs/InAlGaAs QDs grown by molecular beam epitaxy which were integrated into a circular Bragg grating resonator using hyperspectral imaging and deterministic placement. This material platform and resonator design combined have shown to be a source of indistinguishable single photons with reduced lifetime due to the Purcell enhancement \cite{Kaupp.2023, Kim.2025, Hauser.2025}. The deterministic placement of the photonic nanostructure ensures that the QD couples well to the cavity mode and that the polarization bias imprinted by the CBG for displaced emitters is minimized \cite{Peniakov.2024, Buchinger.2025}.

\subsection*{Zeeman splitting and lifetime oscillations}
\label{sec:splittings}

\begin{figure}[t!]
    \centering
    \includegraphics[width=.99\textwidth]{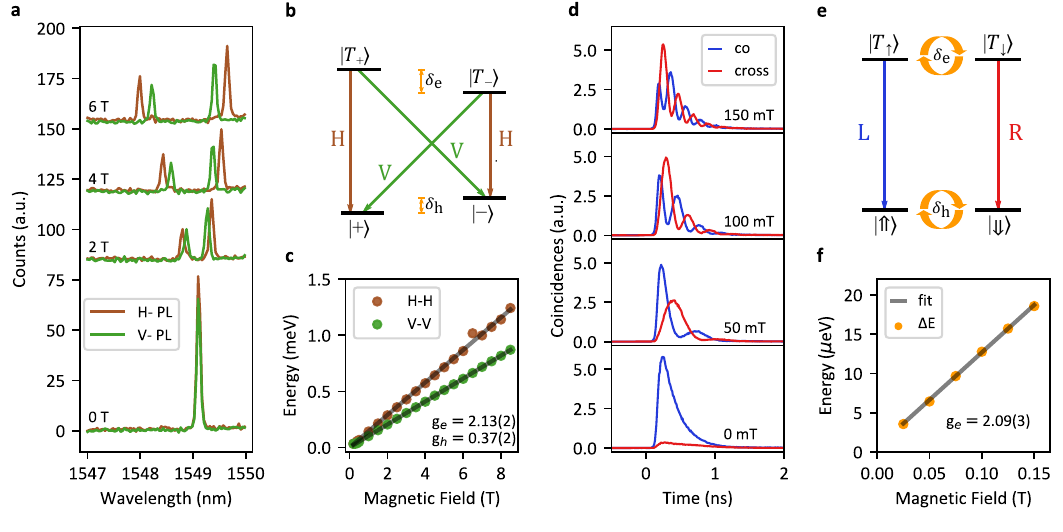}
    \caption{
    \textbf{Splittings and g-factors from polarizations resolved measurements in magnetic field.} \textbf{(a)} Spectral line filtered in H- and V-polarization. The lines split for increasing magnetic field, with a stronger splitting observed for the H-component. \textbf{(b)} Sketch of the trion energy level scheme. When the splitting is sufficiently large, the optical transitions are governed by linearly (H/V) polarized photon emission. \textbf{(c)} Energy splitting between H (brown)- and V (green) polarized emission lines. From the splitting, the g-factors of excited and ground level spin can be calculated. \textbf{(d)} Time- and polarization-resolved trace of the $T^{+}$ emission at zero magnetic field and for small fields $B_x\leq$ \SI{0.15}{T}. For \SI{0}{T}, a polarization memory of \(0.865 \pm 0.001\) can be extracted. For $B_x>$ \SI{0}{T}, the exponential decays of the trion lifetime traces are modulated by oscillations that stem from spin precessions. The frequency is proportional to the induced energy splitting of the excited state, as illustrated in \textbf{(e)}. \textbf{(f)} Linear fit to the observed energy splitting. From this, a g-factor of $2.09 \pm 0.03$ can be calculated.
    } 
    \label{fig:splittings}
\end{figure}

We analyzed the Zeeman splitting of a positive trion in the presence of an in-plane magnetic field. At \SI{0}{\tesla}, the \(X^{+}\) transition shows no resolvable splitting for orthogonal linear polarizations $H$ and $V$, as is depicted in Fig.~\ref{fig:splittings}~(a). It can further be observed that, when ramping up the magnetic field, the trion splits into four distinct lines. The energy of each spectral line marks a transition between eigenstates of the double-$\Lambda$ system formed by the \(X^{+}\) and the hole $h$ in the ground state, as sketched in Fig.~\ref{fig:splittings}~(b). The splitting \(\delta_{h}\) of the hole eigenstates \(\ket{+}\), \(\ket{-}\) is characterized by the hole g-factor \(g_{h}\), while the splitting \(\delta_{e}\) of the trion eigenstates \(\ket{T_{+}}\), \(\ket{T_{-}}\) is governed by the g-factor \(g_{e}\) of the unpaired electron.

We extracted the g-factor of the electron and hole by analyzing the Zeeman splitting shifts of the polarization resolved spectral peaks. The energy separations between the two $H$- and $V$-polarized transitions, $\Delta H$ and $\Delta V$, were fitted linearly as a function of the magnetic field as shown in Fig.~\ref{fig:splittings}~(c). From these fits, we obtained $g_{e,x} = 2.13 \pm 0.02$ and $g_{h,x} = 0.37 \pm 0.02$ following \(\Delta E_{i}=\mu_{B}~g_{i,x}~\Delta B_{x}\) with \(\mu_{B}\) being the Bohr magneton. Consistent with previous work on (InAs/InGaAs) QDs with a comparable host material and strain \cite{Peniakov.2025} we assign the larger g-factor to the unpaired electron in the trion state, while the smaller value corresponds to the hole in the ground state.

In the absence of a magnetic field, the emission lines from a trion are degenerate in energy according to Kramers' theorem and thus not spectrally distinguishable. Furthermore, the following selection rules along the $\hat{z}$ direction apply:
\begin{equation}
\label{eq:selection_rules}
    \ket{T_{\uparrow / \downarrow}} \longmapsto \ket{{\Uparrow} / {\Downarrow}} \ket{{L}/{R}},
\end{equation}
with the trion states $\ket{T_{\uparrow / \downarrow}}$, ground state hole spin states $\ket{{\Uparrow} / {\Downarrow}}$ and photon polarizations states $\ket{{L}/{R}}$. This mapping between spin states and photon polarization states leads to an experimentally observable polarization memory effect: Exciting the trion (quasi-) resonantly with $R$~($L$) circularly polarized light is, upon recombination, followed by the emission of a photon with the same polarization $R$~($L$).

Here, we applied a quasi-resonant excitation scheme that begins with a single hole, denoted \(\ket{h}\), confined in the QD. Upon optical injection of an electron-hole pair, the system is excited to the higher-energy trion state \(\ket{T^*}\) from where it transitions to the trion ground level \(\ket{T}\) via phonon-assisted relaxation. This scheme has been shown to enable writing of a spin qubit \cite{Peniakov.2025} and was furthermore used for cluster state generation in the NIR range \cite{Cogan.2023}. We selected a resonance by performing a photoluminescence-excitation (PLE) measurement. For this, a tunable cw-laser was scanned from \SIrange{1465}{1540}{\nano \meter} and a resonance at \(\sim \SI{1521}{\nano \meter}\) was chosen (cf. supplement section \ref{sec:app_PLE}). Fig.~\ref{fig:splittings}~(d, bottom panel) depicts a measurement at \SI{0}{\tesla} from which we could infer a degree of circular polarization (DOCP) of \((86.5 \pm 0.1) \%\). This value is approaching values obtained in the NIR range using LA-phonon assisted excitation schemes (\(\approx 95 \%\) \cite{Coste.2023}). The observed deviation from ideal spin-retaining excitation likely reflects contributions from residual scattering that does not preserve the spin as well as a minor misalignment between the QD and CBG positions. Throughout the rest of this work, we utilize this polarization memory to initialize the trion in a state $\ket{T_{\uparrow}}$ ($\ket{T_{\downarrow}}$) using $L$~($R$)-polarized laser illumination. 

The splitting of the trion energy levels in a weak magnetic field ($B_{x}\leq$ \SI{150}{mT}) is not strong enough to separate the emission lines spectrally. In this regime, the effect of the magnetic field is as follows: Since $\ket{T_{\uparrow}}$ and $\ket{T_{\downarrow}}$ are both superpositions of the trion eigenstates $\ket{T_{+}}$ and $\ket{T_{-}}$ and $B_{x}$ induces a splitting $\delta_{e}=\mu_{B}g_{e}B_{x}$ of those states, a state $\ket{T_{\uparrow}}$ will evolve to the state $\ket{T_{\downarrow}}$ in a time $T={h}/{2\delta_{e}}$, with $h$ being Planck's constant (cf. Fig.~\ref{fig:splittings}~(e)).

This relation becomes plainly visible when measuring the time trace of the trion with excitation and detection in the $R$/$L$ polarization basis. Fig.~\ref{fig:splittings}~(d) depicts the time trace for magnetic fields ranging from \SI{0}{T} to \SI{150}{mT}. For non-zero magnetic fields, the exponential decay is modulated by oscillations. We fitted the time traces to extract the oscillation frequencies and converted them into energies. The resulting $\Delta E (B)$ dataset is displayed in Fig.~\ref{fig:splittings}~(f). This enabled us to determine the g-factor from the slope of a linear fit to $g_e=2.09\pm0.03$. Within a \(2\sigma\) interval, this value is in agreement with $g_e$ obtained above for higher magnetic fields.

This leaves the remaining hole g-factor, $g_h$, to govern the splitting of the ground-state eigenstates. Analogous to the case of the excited state, a hole spin $\ket{\Uparrow}$ precesses in time around the eigenstate axis. Our observation of this precession is discussed in the following sections.

\FloatBarrier

\subsection*{Ground state spin precessions}
\label{sec:spin_decoherence}

In contrast to the measurements presented in the previous section, the ground-state spin dynamics can not be directly monitored via the time trace of the emitted photon. Here, we made use of the selection rules along the $\hat{z}$-direction again. By performing two-photon correlation experiments we could herald the ground state spin in either $\ket{\Downarrow}$ or $\ket{\Uparrow}$ by detecting a $R$- or $L$-polarized photon and thus study the evolution of the state before detection of the second photon. In a first experiment we studied the spin under continuous-wave quasi-resonant excitation as has been demonstrated in recent years by various groups \cite{Coste.2023, Serov.2025}. Following this, we extend the results by following a different approach that utilizes excitation with two pulses with variable delay and polarization.

\subsubsection*{Autocorrelation measurement with continuous excitation}
\label{sec:cw_g2}

\begin{figure}[t!]
    \centering
    \includegraphics[width=.99\textwidth]{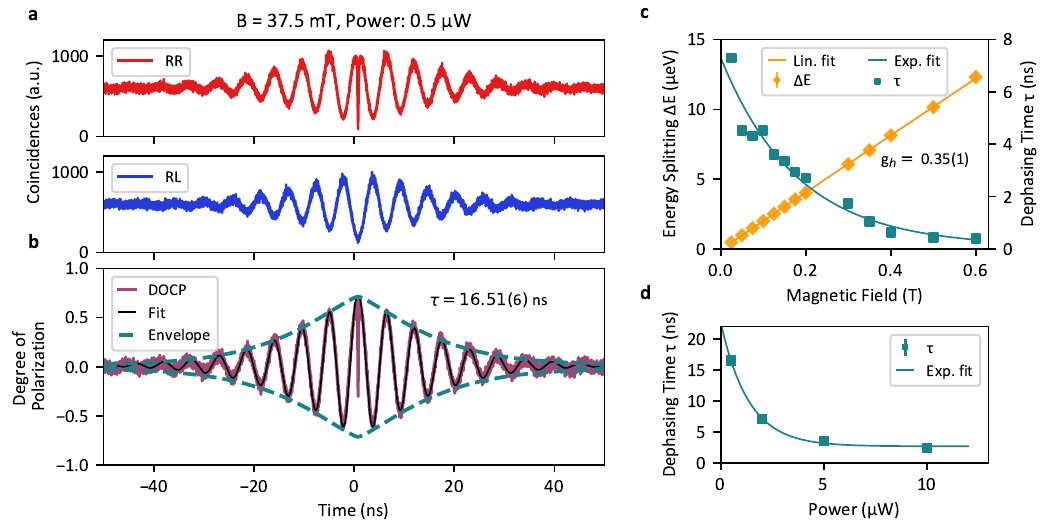}
    \caption{
    \textbf{Polarization resolved autocorrelation measurement.}
    The trion was excited continuously at different magnetic fields and with varying laser powers. The polarizations were set to either R/R (excitation/detection) or R/L, enabling the observation of ground state spin precessions. \textbf{(a)} R/R (top, red) and R/L (bottom, blue) time traces for \SI{0.5}{\micro \watt} and $B=$~\SI{37.5}{mT}. Calculating $(RR-RL)/(RR+RL)$ yields the DOCP displayed in \textbf{(b)}. The decaying envelope used to modulate the fitted cosine function yields a coherence time of $\tau=(16.51\pm0.06)$ \SI{}{ns}. \textbf{(c)} Oscillation frequency~/~energy splitting extracted at various magnetic fields. From the energy splittings, the g-factor can be determined to $0.35 \pm 0.01$. Additionally, one can observe a decrease in \(\tau\) for increasing magnetic fields. \textbf{(d)} Laser-power dependent dephasing time of the hole spin. Similar to increasing magnetic fields, \(\tau\) follows an exponential decay for increasing power.
    }
    \label{fig:cw_g2}
\end{figure}

We performed the measurements utilizing a Hanbury Brown-Twiss-like setup configuration, with cw-excitation and polarization projection before the beam splitter. 
Fig.~\ref{fig:cw_g2}~(a) displays the autocorrelation data measured in parallel (top, R/R) and orthogonal (bottom, R/L) polarization configuration (excitation/detection), with a laser power of \SI{0.5}{\micro W} and an in-plane magnetic field of \SI{37.5}{mT}. In R/R configuration, there is an increased possibility that the trion is excited back into the $\ket{T_{\downarrow}}$ state after it's decay into the $\ket{\Downarrow}$ state. Note that, for time delays close to $t=0$, the photon antibunching, which is a benchmark for single photon sources, causes a sharp dip in the signal.
On the other side, when an $L$-polarized photon is detected - heralding the spin in the $\ket{\Uparrow}$ state - the hole spin has to precess half a period ($T_h/2$) before efficient re-excitation can take place. This causes a $\pi$-shift in the observed oscillations. Fig.~\ref{fig:cw_g2}~(b) displays the DOCP, calculated according to $(RR-RL)/(RR+RL)$ from the data of panel (a) and fitted by a combination of cosine function \(cos(2\pi ft+\phi)\) and stretched decaying envelope term \(e^{-(t/\tau)^\alpha}\). This enabled us to extract a dephasing time of $\tau=(16.51\pm0.06)$~\SI{}{\nano \second} for the ground state spin precession. 

We studied the dependence of the fit parameters $f$, $\tau$ and $\alpha$ on both the magnitude of the magnetic field and the excitation power. Parallel to the excited state evolution described in the previous section, the frequency $f$ is governed by the g-factor and follows $f=\delta_{h} /h$ with $\delta_{h}=\mu_{B}g_{h}B_{x}$. Fig.~\ref{fig:cw_g2}~(c) features a linear fit to $\Delta E (B)$. The g-factor resulting from the slope of the fit amounts to $0.35 \pm 0.01$, which is consistent with $g_{h}=0.37 \pm 0.02$ measured earlier within one standard deviation. This confirms that the observed signal undulations stem from the \textit{ground state} spin precessions, as opposed to the oscillations in the lifetime measurements that could be attributed to $g_{e}$. We would like to draw the readers attention towards the magnetic field dependent dephasing time $\tau (B)$ which is also included in Fig.~\ref{fig:cw_g2}~(c). While increasing \(B_{x}\) leads to an increased level splitting and oscillation frequency, the extracted coherence gets clearly reduced. 

A similar dependency of the $\tau$ could be identified with regard to the excitation power. As depicted in Fig.~\ref{fig:cw_g2}~(d), repeating the experiment for different excitation powers enabled us to fit an exponential dependence between laser power and dephasing time of the hole spin. Under highly-driven conditions the extracted $\tau$ is significantly reduced which we attribute to a strongly increased re-excitation probability. 

For both studies presented in Fig.~\ref{fig:cw_g2}, $\alpha$ was fitted as a free parameter and found to exhibit values ranging from \(\sim 1.3\) to \(\sim 0.5\) (cf. supplementary material). The extrapolation of $\tau$ towards even smaller laser powers shows that the measured coherence of $\sim$\SI{16}{\nano \second} could be exceeded for an unperturbed system. Therefore, we studied the hole spin under pulsed excitation, which we elaborate in the next section.

\FloatBarrier

\subsubsection*{Probing the spin with pulses}
\label{sec:pulsed_decoherence}

We studied the spin evolution of the ground-state hole using two-photon correlations (2PC). For this measurement, the laser pulse was split in two paths, with the delay in one arm controlled by a motorized linear stage. The delayed pulse is hereafter referred to as \textit{pulse 2} whereas the other pulse is \textit{pulse 1}. For this measurement, we used two separate excitation paths, as well as two detection channels (CH1 and CH2). The detection channels are synchronized to the laser trigger that is created by pulse 1. Both, the excitation polarization and the detection polarization projection can be controlled individually for each pulse and detection channel. Thus, we can excite and project each photon in an arbitrary basis.  The in-plane magnetic field was set to $B_{x}=$ \SI{150}{\milli \tesla}.

\begin{figure}[t!]
    \centering
    \includegraphics[width=.99\textwidth]{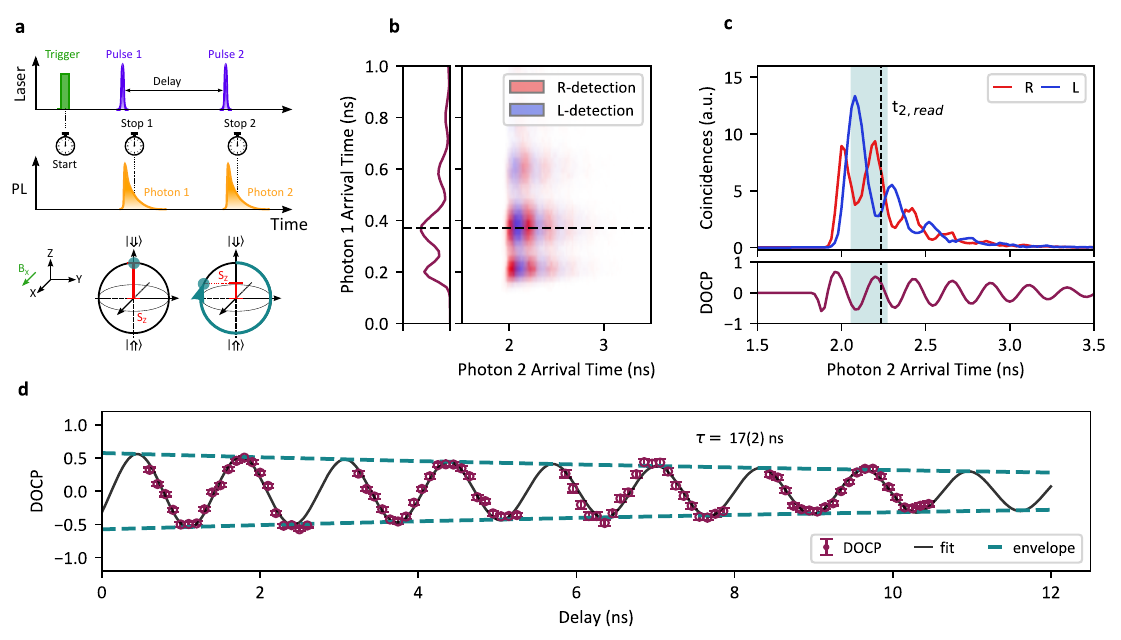}
    \caption{
    \textbf{Pulsed coherence time measurement.} 
    The hole spin is probed utilizing a two-photon-correlation experiment. \textbf{(a)} Schematic illustration of the measurement sequence. The experiment is synchronized by an electronic trigger signal from the laser which starts the measurement clock. The emitted laser pulse is split in two pulses and the second pulse arrives delayed due to an optical path of controllable length. The polarization for both pulses can be controlled individually. The same applies for the projection of the resulting photons, whose arrival in two separate channels CH1 and CH2 stops the clock. Detecting the first photon in $R$ (after $R$-excitation) heralds the spin in $\ket{{\Downarrow}}$. After undisturbed precession with the Larmor frequency around an in-plane magnetic field of \SI{150}{\milli \tesla}, the hole spin is probed with an $H$-polarized laser pulse and projected to both $R$ and $L$. \textbf{(b)} Resulting two-dimensional time trace of photon 2 for a pulse delay of \SI{1.6}{\nano \second}. The data of both circular polarizations is superimposed in a colormap. The time traces are analyzed along the horizontal direction with regard to a fixed arrival time of the CH1 photon at $t_{X,CH1}=$ \SI{370}{\pico \second} (cf. dashed black line). The resulting one-dimensional time traces are displayed in \textbf{(c)}. Top panel: Correlated $\langle RR \rangle$ (red) and $\langle RL \rangle$ (blue) signals. A phase shift of $\pi$ can be observed between the oscillating time traces. Bottom panel: Fit of the DOCP resulting from the datasets in the top panel. A vertical dashed line marks the point of readout, which is varied over a window of \SI{220}{\pico \second} (shaded area). \textbf{(d)} DOCP at $t_{2, read}(\Delta t)$. The resulting oscillation reflects the ground-state spin-precession of the hole. From a fit, a precession frequency of $763 \pm 1$ \SI{}{\mega \hertz} and spin dephasing time of $17.0 \pm 2.2$ \SI{}{ns} are extracted.
    }
    \label{fig:pulsed_coherence}
\end{figure}

Fig.~\ref{fig:pulsed_coherence}(a) illustrates the measurement protocol, which is described in detail below. After excitation by pulse 1 ($R$-polarization) and subsequent detection of a $R$-polarized photon in CH1, the spin is heralded in $\ket{\Downarrow}$ according to $\ket{T_{\downarrow}} \longmapsto \ket{{\Downarrow}} \ket{{R}}$. Following this, the hole spin precesses freely around the eigenstate axis in the time $\Delta t$ between the two pulses with a precession time of \(T_{h}(150\text{ mT})=\)~\SI{1.36}{\nano \second}. The $H$-polarized pulse 2 then transfers the spin superposition to the excited state while maintaining its phase. For several delays $\Delta t=$\SIrange{0.61}{10.45}{\nano \second} a pair of two datasets was acquired, projecting the second photon in CH2 to $R$ and $L$ respectively. This enabled us to determine the DOCP after $\Delta t$ to:
\begin{equation}
\label{eq:stokes_parameter}
    DOCP = S_{z}=\frac{{\langle RR \rangle}-{\langle RL \rangle}}{{\langle RR \rangle}+{\langle RL \rangle}},
\end{equation}
with $S_{z}$  being the Stokes parameter in the $Z$-basis.
Fig.~\ref{fig:pulsed_coherence}~(b) maps two overlaid datasets of the second photon arriving in CH2, projected to $R$ and $L$ respectively. Here, the delay of the two laser pulses was set to \SI{1.6}{\nano \second}.
When being summed up and projected on the vertical axis (detection of first photon in CH1), the data show the fingerprint of the oscillating lifetime trace (cf. Fig.~\ref{fig:splittings}~(d)). We analyzed a horizontal cut through the data at $t_{CH1}=370$ \SI{}{\pico \second} as indicated by the black dashed line. This slice of data is plotted in Fig.~\ref{fig:pulsed_coherence}~(c, top) and features counter-oscillating $R/L$ time traces. We would like to highlight that, would the two photons not be measured correlated, exciting with $H$ would not result in an oscillating lifetime. The fact that we observe oscillations in the lifetime here shows that we can effectively initialize the spin with pulse 1 and probe it with the second laser pulse.

We proceed by extracting the DOCP for different delays in the following way:  We chose an arbitrary time \(t_2\) after the excitation time of the second photon \( t_{pulse~2}\), for which we read out \(S_{z}(t_{2,~read})\) with \(t_{2,~read} = t_{pulse~2}+t_{2}\).

Fig.~\ref{fig:pulsed_coherence}~(d) shows the extracted DOCP with respect to the pulse delay for $t_{2}=$~\SI{230}{\pico \second} (cf. dashed vertical line in Fig.~\ref{fig:pulsed_coherence}~(c)). A fit to the data according to
\begin{equation}
\label{eq:fit_function_pulsed_docp}
    A(t)=C+A_0 \exp({-({t}/{T_2^*})^{\alpha}}) \cos{(\omega t + \varphi)}, 
\end{equation}
with \(\omega=2 \pi f\) and \(\alpha = 1\) yields a spin dephasing time of $T_2^*(230~\text{ps})= (17.0 \pm 2.2)$~\SI{}{ns}. Since the fitted value of $T_2^*$ varies slightly with the readout point due to noise on the data, we considered the average over a phase of \(2\pi\) of the excited state precession, see shaded area in Fig.~\ref{fig:pulsed_coherence}~(d) and supplement Fig.~\ref{fig:app_pulsed_decoherence}. This yields a mean dephasing time of $\overline{T}_2^*=(15.9\pm1.7)$~\SI{}{\nano \second}, which is in agreement with $\tau=(16.51\pm0.06)$~\SI{}{\nano \second} obtained for continuous-wave excitation in the previous section. Similarly, the precession frequency is averaged over this window to $\bar{f}=0.761 \pm 0.002$~\SI{}{\giga \hertz} which calculates to a g-factor of $0.362 \pm 0.001$, in agreement with the value $g_{2}$ found earlier in this work (cf.~Fig.~\ref{fig:splittings}). 

%% file: Sections/3_Discussion.tex
\section{Discussion}
\label{sec:discussion}

We proceed by comparing the g-factors measured here with other works on InAs/InAlGaAs QDs grown on a InP substrate. Extensive studies by \textit{Belykh et al.} on ensemble emission of nominally similar dots using pump-probe spectroscopy have found electron g-factors \(\sim 1.9\) and hole g-factors \(\sim 0.6\), with huge out-of-plane and small in-plane anisotropy \cite{Belykh.2015, Belykh.2016}.Compared to these values, the single QD investigated here shows a larger difference \(|g_e - g_h|\) with \(g_e \simeq 2.15\) and \(g_h \simeq 0.35\) both deviating from the above mentioned values by \(0.25\). This deviation is significantly larger than in-plane g-factor spreads of the QD ensemble determined by \textit{Belykh et al.} to \(\Delta g < 0.1\) and therefore probably originates from a slightly different dot shape leading to different confinement, In-Ga intermixing or strain. We interpret our results to strengthen our identification of the studied line to be a positive trion, with the excited state spin dynamics governed by the electron and the ground-state single-spin precession determined by the hole.

In addition, this identification is in agreement with the long coherence time of the precessing hole described in this work. Studies on \(\SI{900}{\nano \meter}\) emitting dots conducted via pump-probe experiments and in the absence of an external magnetic field have revealed an electron spin depolarization time of \(\sim \SI{2}{\nano \second}\) \cite{Bechtold.2015}. Similar reports comparing electron and hole spins have measured one order of magnitude larger hole spin dephasing times of \(\sim \SI{20}{\nano \second}\) which can be attributed to a weaker hyperfine interaction of the heavy-hole with surrounding nuclei \cite{Cogan.2018}. The \(T_{2}^{*}\) values we could extract from our measurements mark a major step towards closing the gap between NIR- and telecom QDs, with record values still residing in the NIR-range \cite{Huthmacher.2018}.

Another novelty factor that our results entail is the approach for measuring \(T_{2}^{*}\) displayed in Fig.~\ref{fig:pulsed_coherence}. We could probe the unperturbed evolution of our hole spin between two laser pulses by \textit{heralding} the spin during detection of the first photon. While measurements that are conceptually similar, like continuous excitation or evaluating oscillations in the trion lifetime trace can also quantify \(T_{2}^{*}\), they fall short in different ways: The first case drives the \(\ket{\Downarrow} \mapsto T_{\downarrow}\) transition continuously which shortens the measured coherence time exponentially with applied laser power (see our results in Fig.~\ref{fig:cw_g2}). The latter case was used by this and other groups \cite{Peniakov.2025, Laccotripes.2025} in order to provide a lower boundary for the hole spin coherence, but is ultimately limited by the radiative lifetime of the trion. Many additional studies have measured \(T_{2}^{*}\) using Ramsey-interference. However, for magnetic fields \(B_{x}^{ext}>\SI{4}{\tesla}\), the pure spin dephasing time has been shown to decrease inversely with the external magnetic field, following \(T_{2}^{*} \propto 1/B_{x}^{ext}\) \cite{Huthmacher.2018}. Since one major application for coherent spin qubits is the generation of linear cluster states, and since for this protocol the in-plane magnetic field is usually in the \SIrange{10}{500}{\milli\tesla} range when using polarization encoding, we believe that it is best to quantify \(T_{2}^{*}\) at similar magnetic fields. At the same time, it is therefore reasonable to assume that there may be an optimal magnetic field \(B_{x}^{opt}\) other than the \SI{150}{\milli\tesla} studied with pulsed excitation here, for which \(T_{2}^{*}\) of our hole spin exceeds \SI{17}{\nano\second} which we plan to investigate further in upcoming experiments.

\input{Sections/4_Conclusion}

%% file: Sections/4_Conclusion.tex
\subsection*{Conclusion}

We have studied a device consisting of a InAs/InAlGaAs quantum dot integrated in a deterministically placed circular Bragg-grating. We have identified a positive trion in a polarization-resolved magnetic field series under above-bandgap illumination and have been able to determine the g-factors for electrons and holes in the QD to $g_{e} = 2.134 \pm 0.016$ and $g_{h} = 0.367 \pm 0.016$. Utilizing a p-shell resonance found in a PLE measurement, we have found a polarization memory of \((86.5 \pm 0.1) \%\) and have studied the ps-resolved time evolutions of excited- and ground state spin of the trion in a weak magnetic field. From the observed oscillation frequencies, we have been able to confirm the earlier found g-factor values. Furthermore, we have quantified the dephasing time of the ground-state hole in a second-order autocorrelation measurement using low excitation powers. We then have tested the validity of this approach by probing the ground-state spin dynamics with ultrashort optical pulses, after heralding the spin in a two-photon correlation experiment. There, we have found a dephasing time of $15.9 \pm 1.7$ \SI{}{ns}, which not only confirms the preceding measurement but, to the best of our knowledge, is the longest pure dephasing time measured for an electron or hole spin in telecom-emitting quantum dots so far.

The positive trion studied in our device exhibits long ground-state spin coherence and short excited state lifetime due to cavity integration. The combination of these features makes it an ideal candidate for studying more complex spin manipulation protocols like dynamical decoupling or for extending to three-photon correlations.

%% file: Sections/5_Methods.tex
\section*{Methods}
\small

\subsection*{QD growth}
\label{sec:methods_QD_growth}

The InAs quantum dots (QDs) were grown by gas-source molecular beam epitaxy (MBE) on an InP substrate to achieve emission in the Telecom C-band. 
The heterostructure consists of an \SI{226}{\nano\meter} InAlGaAs membrane lattice-matched to InP, serving as the optical cavity layer. Below the membrane, an InGaAs etch-stop layer was incorporated.
The QDs were formed via Stranski–Krastanov growth by depositing a few monolayers of InAs under elevated arsenic flux, followed by a brief ripening step and capping with the surrounding InAlGaAs barrier. Digital alloying of the quaternary layers was employed to enhance material homogeneity and control strain. The full growth procedure and structural characterization are detailed in previous work, see Refs.~\cite{Kaupp.2023, Kim.2025, Hauser.2025}.%%, Kohr2025ManuscriptInPreparation}.

\subsection*{CBG design}
\label{sec:methods_CBG_design}

The circular Bragg grating (CBG) cavities were optimized using three-dimensional finite-difference time-domain (FDTD) simulations to achieve strong Purcell enhancement at the targeted Telecom C-band wavelengths. The design parameters, including center disk radius, grating period, and gap width, were systematically varied to maximize the cavity quality factor and minimize the mode volume while maintaining efficient vertical out-coupling. The resulting geometry supports a Gaussian far-field profile and robust performance against fabrication deviations, following the design principles established in previous works \cite{Kaupp.2023, Kim.2025, Hauser.2025}.

For fabrication, the optimized CBG pattern was defined by high-resolution electron-beam lithography and transferred into the InAlGaAs membrane by inductively coupled plasma etching using an \(\mathrm{Ar/Cl_2}\) chemistry. Prior to this, a \SI{480}{\nano\meter} dielectric AlOx layer was deposited by sputtering to form the bottom spacer layer, followed by an evaporated Au mirror providing high reflectivity and vertical mode confinement. The sample was then bonded upside-down onto a GaAs carrier substrate using a benzocyclobutene (BCB) adhesive in a flip-chip process, ensuring thermo-mechanical stability. The original InP substrate and buffer layers were subsequently removed by selective wet etching, exposing the InAlGaAs membrane containing the quantum dots. The final device stack thus comprises the suspended InAlGaAs membrane with integrated QDs, the sputtered AlOx layer, and the gold mirror, forming a high-performance CBG cavity suitable for deterministic QD coupling.

\subsection*{Deterministic CBG placement}
\label{sec:methods_imaging}

In order to center the optimized CBG as good as possible around a preselected QD, we used a hyperspectral imaging technique combined with two image photoluminescence imaging.

As a first step, alignment markers with a periodic marker design were deposited on the sample surface, dividing it into different fields of \SI{37}{\micro\meter} by \SI{37}{\micro\meter}. An excitation laser was focused in the back-focal plain of the objective, thus illuminating the whole field homogeneously. A 4-f setup in combination with a Czerny-Turner spectrometer (slit closed to \SI{50}{\micro \meter}) was used to analyze a strip of the real space image spectrally. Moving /Scanning the lens in-front of the spectrometer sideways shifts the image of the sample in respect to the entrance slit, thus changing the analyzed strip. This yields the hyperspectral information of the whole field. From the resulting 3D data cube, suitable QDs were selected for further processing.

To provide the most accurate spatial information, a second camera was utilized together with LED illumination. In the photoluminescence imaging step two frames were acquired immediately after each other: one with laser illumination and no LED and one frame vice versa. Overlaying both frames shows the QD emission (without spectral information) with respect to the periodic markers. From a fit of both features, the position of the QD with respect to the markers could be determined and translated into coordinates for CBG fabrication by the EBL system. For more detailed information see \textit{Buchinger et al.} \cite{Buchinger.2025}.

\subsection*{Excitation lasers}
\label{sec:methods_excitation_lasers}

For the experiments presented in this work three different lasers were used. For the polarization resolved measurements displayed in Fig.~\ref{fig:splittings}~(a-c) a laser diode with an emission wavelength of \SI{1064}{\nano \meter} was used. The PLE scans as well as results of Fig.~\ref{fig:cw_g2} were obtained through a continuously tunable, fiber coupled cw-laser (Toptica\(^{TD}\) CTL 1500). The laser pulses required for the results presented in Fig.~\ref{fig:splittings}~(d-e) and Fig.~\ref{fig:pulsed_coherence} were produced by a wavelength tunable, picosecond OPO-laser with \SI{80}{\mega \hertz} repetition rate (APE picoEmerald\(^{TD}\)).

In order to produce two pulses with an adjustable delay we divided the laser with a fiber beam splitter into two paths. The first path remains in fiber and is guided directly to the setup. The second path is coupled into free space and routed to a dedicated delay setup. There, a motorized linear translation stage with a travel range of \SI{100}{\milli \meter} controls the optical path length before coupling the laser pulse back into fiber. To extend the adjustable pulse delay, the beam is reflected multiple times from the translation stage.

\subsection*{Polarization control and calibration}
\label{sec:methods_polarization_calibration}

We use both, liquid-crystal variable retarders (LCVRs) as well as half-wave plates (HWPs) and quarter-wave plates (QWPs) to translate the excitation and detection polarization from the the laboratory frame to the QD-frame and vice versa. The detection polarization is especially affected by skewing since the emitted PL is first coupled into fiber and then projected in a dedicated setup on the optical table. 

We calibrated the detection polarization through the spectral and temporal signature of the trion as follows: The magnetic field was increased until the emitted PL was distributed between four clearly separable lines (at \(\sim \SI{4}{\tesla}\)). The polarization was then rotated such that detecting \(H\) yields the highest- and lowest-energy lines whereas detecting \(V\) yields the two middle lines with lower splitting \(V-V\) compared to the \(H-H\) splitting (cf. Fig.~\ref{fig:splittings}~(a-c)). In a next step, we reduced the magnetic field to \(\sim \SI{100}{\milli \tesla}\) and optimized the visibility of the trion time trace oscillations (cf. Fig.~\ref{fig:splittings}~(d)). To do so, we rotated the polarization frame around the \(H/V\) basis until the contrast of the oscillations was maximized (minimized) in the \(R/L\) (\(D/A\)) basis (cf. supplement figure.~\ref{fig:app_pol_correction}). The procedure described in this paragraph was repeated iteratively until the best possible calibration is achieved.

The calibration of the excitation polarization follows a similar routine: We used the oscillating lifetime in a weak magnetic field to optimize on the contrast in the \(R/L\) basis; excitation in every other basis (\(H/V\) or \(D/A\)) should result in a flat time trace. The right \(H/V\) basis for the excitation was found through polarization resolved PLE measurements in high magnetic fields.

\normalsize

%% file: Sections/6.1_Acknowledgements.tex
\section*{Acknowledgements}

\subsection*{Funding}
The authors acknowledge the support of the state of Bavaria and the German Federal Ministry of Research, Technology and Space (BMFTR) within Project PhotonQ (FKZ: 13N15759) and QR.N (FKZ: 16KIS2209) as well as QuNET+ICLink (FKZ: 16KIS1975). Reza Hekmati, Mohamed Helal and Tobias Huber-Loyola acknowledge financial support from the BMFTR within the Project Qecs (FKZ: 13N16272). The authors are furthermore grateful for the support by the National Research Foundation (NRF) of South Korea within the international joint research program on \textit{Chip-scale Scalable Quantum Light Sources and Photonic Integrated Circuit Technology Development} (RS-2023-00284018).

% \subsection*{Author contributions}
% J.M.M, R.H., M.H. and G.P. built the setup and conducted the experiments. J.M.M and R.H. analyzed the data and wrote the manuscript. J.K., Y.R., J.K. and Q.B. fabricated and pre-characterized the sample. S.H., A.T.P. and T.H.-L. guided the work and acquired funding.

%% file: Sections/6.2_Declarations.tex
\section*{Declarations}

\subsection*{Competing interests}
The authors declare that they have no competing interests.

\clearpage

%% file: Sections/Supplement.tex
\nolinenumbers
\clearpage
\appendix
\vspace*{1cm} % space after the title before content starts
\begin{center}
    {\LARGE \bfseries Supplementary Material}\\[0.5em]
    \rule{0.5\textwidth}{0.4pt}
\end{center}
\vspace{2cm} % space after the title before content starts
\setcounter{section}{0}
\setcounter{page}{1}
\renewcommand\thesubsection{S\arabic{subsection}}
\setcounter{figure}{0}
\renewcommand\thefigure{S\arabic{figure}}
\renewcommand\thetable{S\arabic{table}}

%%% START ENTRIES

\subsection{Device and spectrum}

\begin{figure}[htb!]
    \centering
    \includegraphics[width=.80\textwidth]{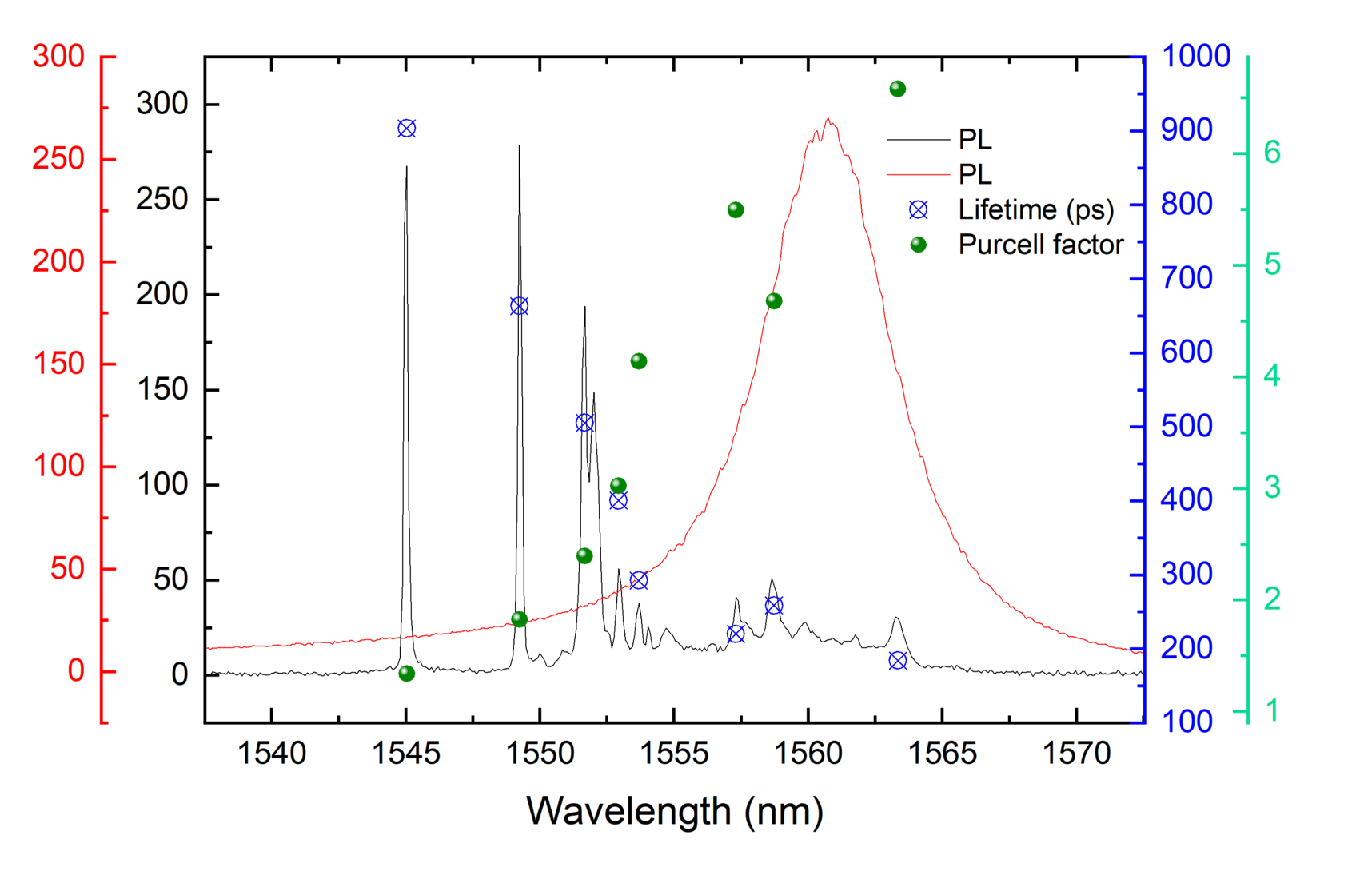}
    \caption{
    \textbf{$\mu$-PL spectrum and cavity enhanced radiative lifetime measurements.} 
    Black: Spectrum under above-bandgap laser illumination. Two prominent emission lines appear below \SI{1550}{nm}. Red: Cavity emission under strong laser pumping. The cavity is red-shifted around \SI{10}{nm} from the targeted center wavelength. Blue: Measurements of the radiative lifetime at the position of prominent PL-features. The lifetime significantly decreases for lines that couple to the cavity. Green: Purcell factor calculated from the decreased radiative decay times. The Purcell enhancement follows the cavity's spectral shape.
    }
    \label{fig:app_cavity}
\end{figure}

Fig.~\ref{fig:app_cavity} shows the spectrum of the QD studied in this work under above-bandgap excitation for small excitation powers (black, single lines) as well as for high excitation powers (red). The latter spectrum shows the resonance of the CBG cavity.

\FloatBarrier
\newpage

\subsection{Identification of exciton and trion}
\label{sec:app_power_pol_series}

\begin{figure}[htb!]
    \centering
    \includegraphics[width=.99\textwidth]{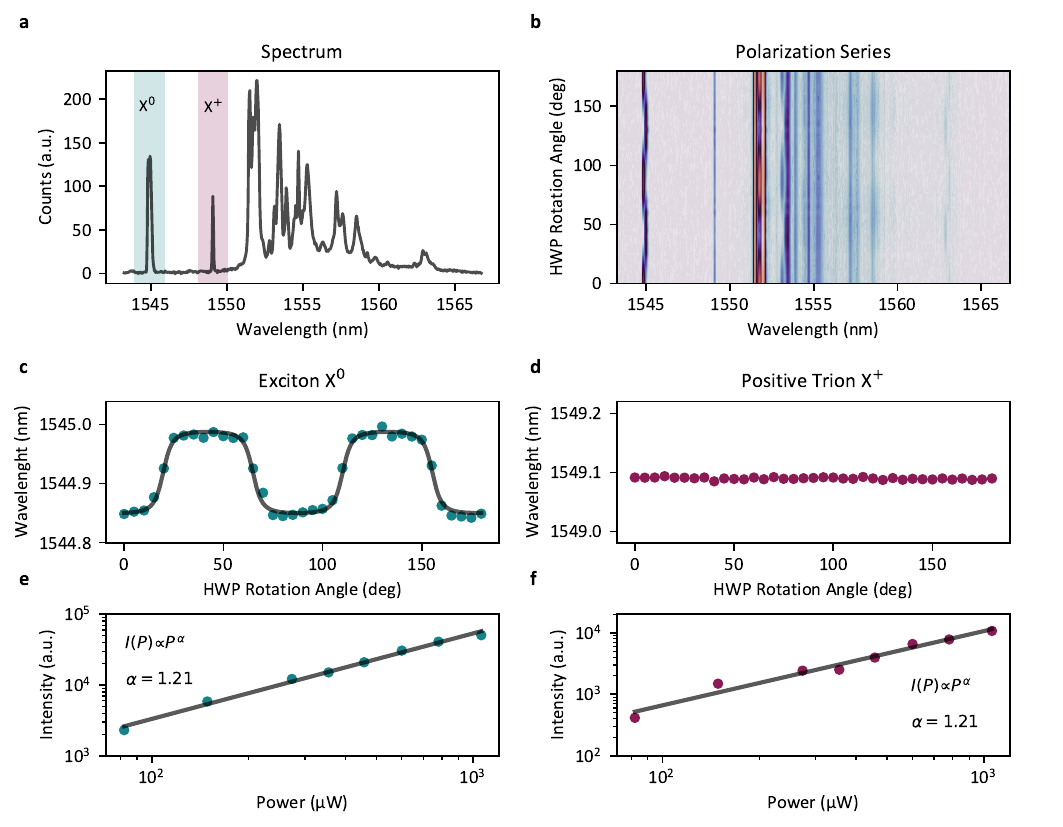}
    \caption{
    \textbf{Analysis of peak position and intensity under varying excitation powers and detection polarizations.} 
    (a) QD spectrum with the exciton and trion line highlighted. (b) Colormap constructed from all spectra of a polarization series, with the HWP angle in the detection tuned from \SIrange{0}{180}{\degree}. (c-d) Peak position of \(X^{0}\) and \(X^{+}\) from polarization series. The exciton shows a fine-structure-splitting, whereas the trion wavelength remains polarization-independent. (e-f) Respective power dependent peak intensities. Both fitted lines showcase a power-law coefficient of \(\sim 1.2\). 
    }
    \label{fig:app_power_pol_series}
\end{figure}

The positive trion studied in this work is located at \(\sim \SI{1549}{\nano\meter}\) (cf. Fig.~\ref{fig:app_power_pol_series}~(a)). We could identify the exciton and trion in our sample by means of power- and polarization series using above-bandgap continuous-wave (cw) excitation. One hallmark of a trion is the lack of any observable fine-structure splitting (FSS) which can become apparent when rotating the polarization angle in the detection. While the neighboring exciton spectral line showed a clear FSS of \(\sim~\)\SI{71}{\micro \electronvolt}, the trion line investigated here showed no splitting as is depicted in Fig.~\ref{fig:app_power_pol_series}~(b-d). For both, \(X^{0}\) and \(X^{+}\), we furthermore measured a power-law coefficient of \(\alpha \sim 1.2\) according to \(I(P) \propto P^{\alpha}\).

\FloatBarrier
\newpage

\subsection{PLE data}
\label{sec:app_PLE}

\begin{figure}[hbt!]
    \centering
    \includegraphics[width=.99\textwidth]{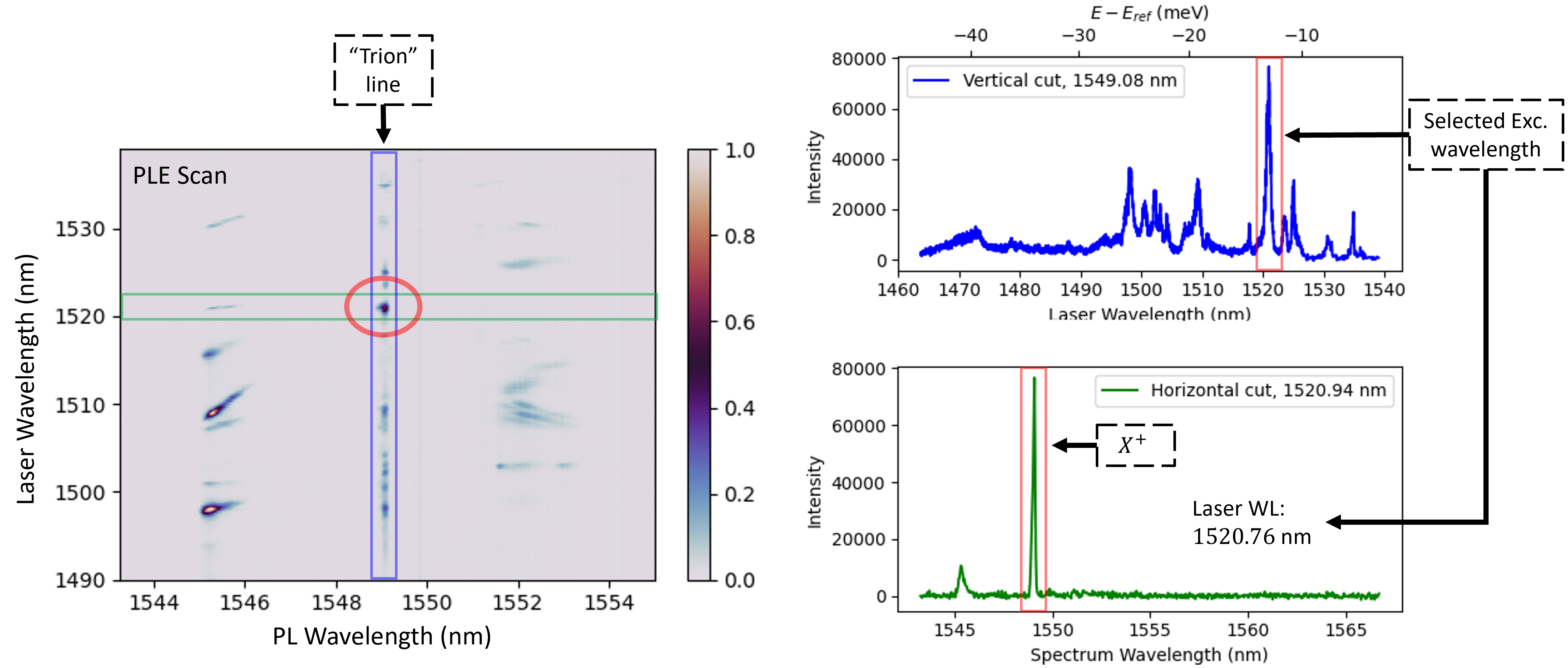}
    \caption{
    \textbf{PLE scan.}
    Left side: Colormap showcasing the resonances of the QD spectral emission lines for the scanned excitation laser wavelengths.
    Right side: Vertical (blue) and horizontal (green) cuts through a resonance found at \(\lambda_{exc}\approx\)\SI{1521}{\nano \meter}.
    }
    \label{fig:app_PLE}
\end{figure}

The exciting laser was scanned from \SI{1465}{\nano \meter} to \SI{1540}{\nano \meter} and a spectrum of the QD was acquired for each excitation wavelength, producing the colormap on the left side of Fig.~\ref{fig:app_PLE}. A vertical cut at the trion wavelength \(\lambda=\)~\SI{1549.08}{\nano \meter} (blue plot) reveals a bright resonance at \(\lambda_{exc} =\)~\SI{1520.76}{\nano \meter}.
\vspace*{1cm}

\newpage

\subsection{Polarization correction with waveplates}

\begin{figure}[hbt!]
    \centering
    \includegraphics[width=.99\textwidth]{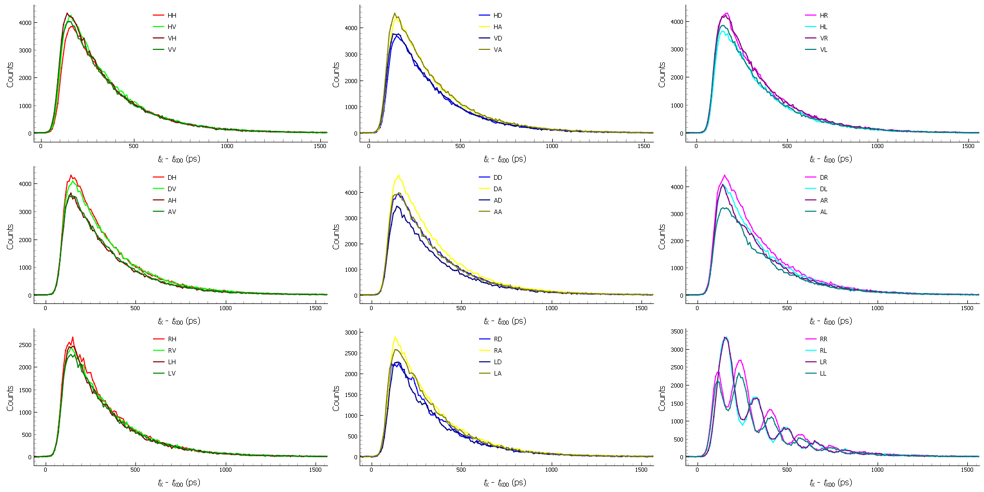}
    \caption{
    \textbf{Optimization on lifetime oscillations.}
    The labeling of each subset of data follows the convention \textit{excitation-polarization/detection-polarization}, so \(RL\) e.g. means a excitation with $R$ polarized laser and a detection of $L$ polarized PL. 
    }
    \label{fig:app_pol_correction}
\end{figure}

Details on the polarization correction and calibration using waveplates is provided in the methods section. Fig.~\ref{fig:app_pol_correction} displays the results of one calibration which was performed iteratively until oscillations in the time trace could only be observed when exciting and detecting in the $R/L$ polarization basis.
\vspace*{1cm}

\newpage

\subsection{Extended datasets on two-photon correlation experiments}

\subsubsection{Power dependent spin dephasing time}
\label{sec:app_cw_g2_power_series}

\begin{figure}[hb!]
    \centering
    \includegraphics[width=.99\textwidth]{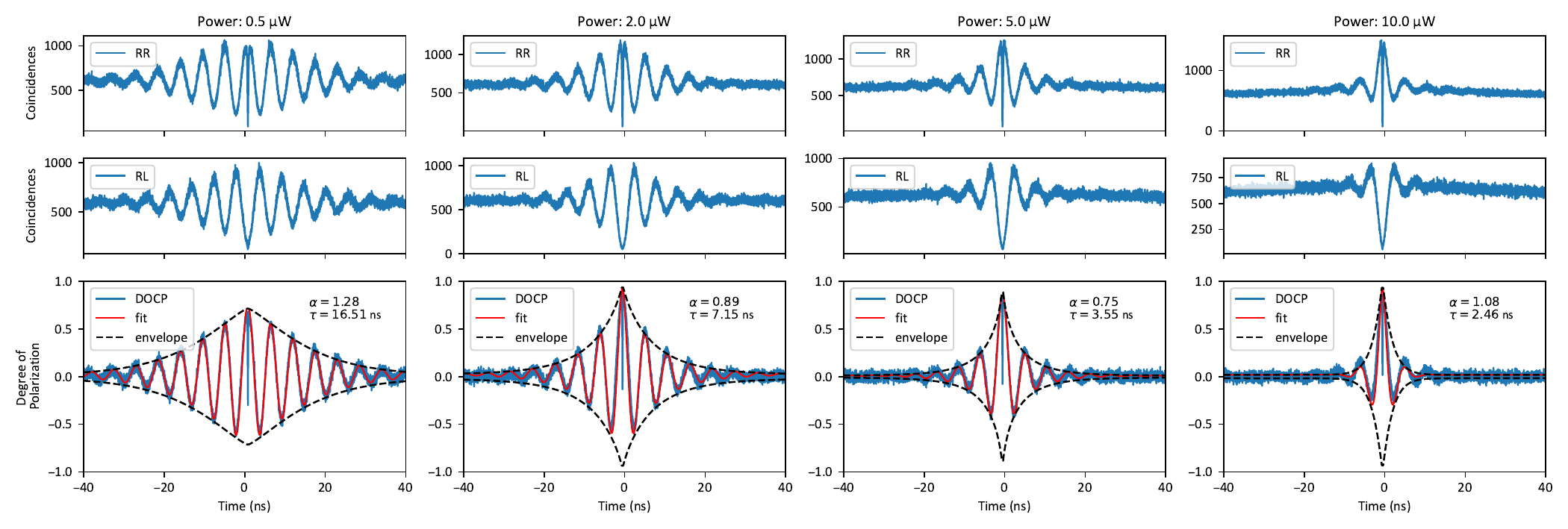}
    \caption{
    \textbf{Power series of cw-g2.}
    The power is increased from the left column to the right column. For each power, $RR$ and $RL$ datasets were acquired and the DOCP was calculated and fitted, extracting the dephasing time from the exponentially decaying envelope.
    }
    \label{fig:app_cw_g2_power_series}
\end{figure}

The data from the power series presented in Fig.~\ref{fig:app_cw_g2_power_series} are fitted using the following function:

\[DOCP(t) = C + A~e^{-(\frac{|t-t_{0}|}{\tau})^\alpha} \cos{(2 \pi f t + \varphi)}\].

A \SI{300}{\pico \second} window around around \(t_0\) is omitted from the fit to exclude the antibunching dip which simplifies the fitting. Four sets of data are analyzed, with \(P=\)~\SI{0.5}{\micro \watt}, \SI{2.0}{\micro \watt}, \SI{5.0}{\micro \watt}, \SI{10.0}{\micro \watt}. 
Below table lists the found values for selected fit parameters:

\begin{table}[h]
    \centering
    \begin{tabular}{c|c|c|c}
        Power [\SI{}{\micro \watt}] & A & $\tau$ & $\alpha$ \\
        \hline \hline
        0.5 & 0.7 & 16.51 & 1.278\\
        2.0 & 0.9455  & 7.148  & 0.8878\\
        5.0 & 0.9576 & 3.548  & 0.7531\\
        10.0 & 0.9635 & 2.463 & 1.078\\
    \end{tabular}
    \caption{Fit parameters for power series of cw-g2 as shown in Fig.~\ref{fig:app_cw_g2_power_series}.}
    \label{tab:placeholder}
\end{table}

\newpage
\subsubsection{Magnetic field dependent spin dephasing time}
\label{sec:app_cw_g2_bfield_series}

\begin{figure}[hb!]
    \centering
    \includegraphics[width=.99\textwidth]{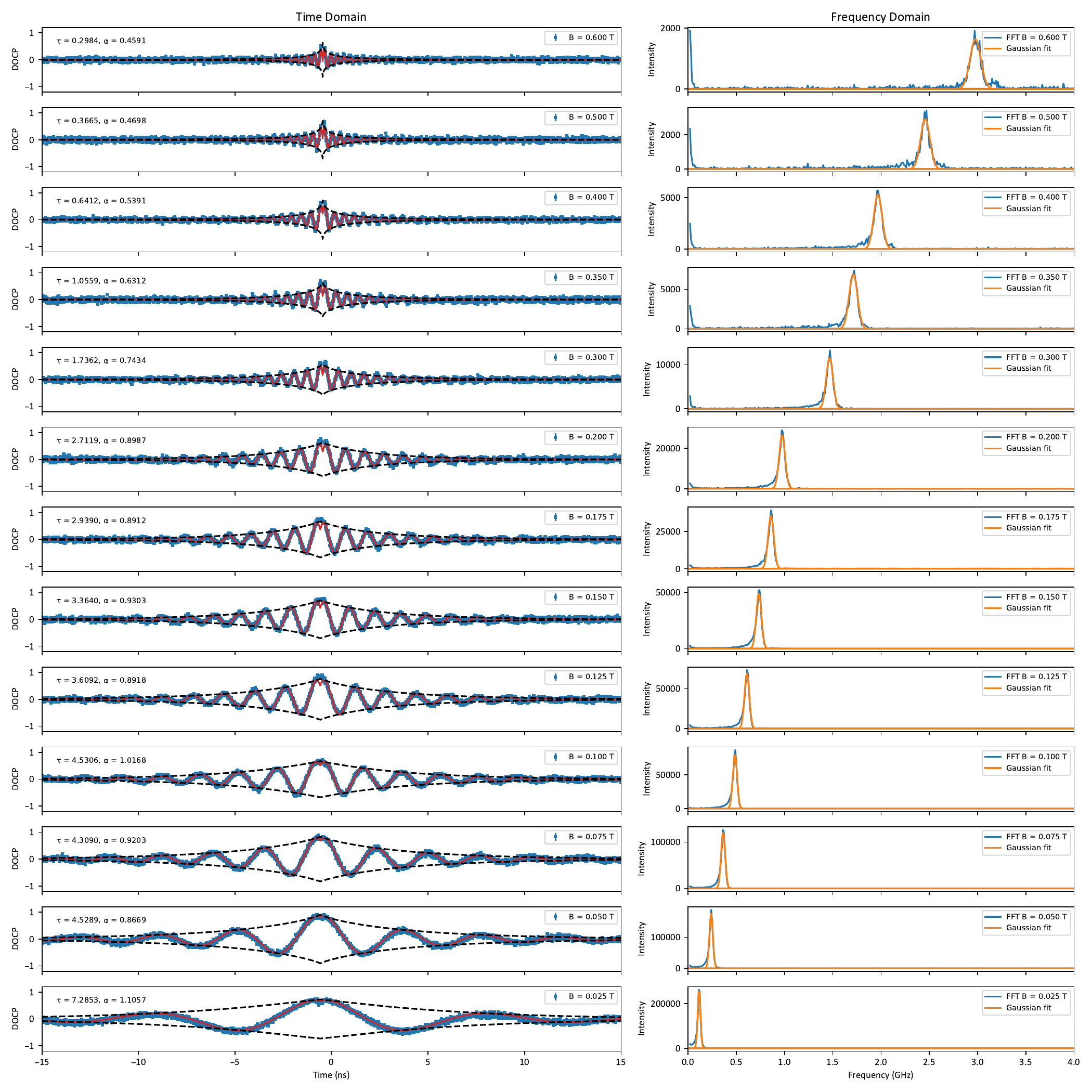}
    \caption{
    \textbf{Magnetic field series of cw-g2.}
    Left column: Fits of the oscillating DOCP, depending on magnetic field. From the fits, the dephasing time \(\tau\) and the parameter \(\alpha\) are extracted. Both are reduced significantly for higher fields. Right column: Fourier transformation of the signals from the left column. The oscillation frequency is determined from Gaussian fits in the frequency domain.
    }
    \label{fig:app_cw_g2_bfield_series}
\end{figure}

The measured DOCPs for increasing magnetic fields are analyzed using the same fitting routine as in the previous section~\ref{sec:app_cw_g2_power_series}. Both dephasing time \(\tau\) and stretching parameter \(\alpha\) are extracted from the fit and displayed in the left subpanels of Fig.~\ref{fig:app_cw_g2_bfield_series} for each magnetic field respectively. Additionally, the frequency was determined from peak-fits after a Fourier transformation.

\newpage
\subsubsection{Discussion of extracted parameter alpha}

In addition to the quasi-static Overhauser field, which has an approximately Gaussian distribution \cite{Merkulov.2002}, there are several other noise sources that the spin qubits in QDs can couple to, including phonons, charge fluctuations, and electrical noise \cite{DeGreve.2011, Prechtel.2016}. Because a hole - in contrast to an electron - couples only weakly to the nuclear bath, these non-nuclear channels can significantly influence the envelope of the dephasing. As a result, fixing \(\alpha = 2\) would implicitly assume that nuclear quasi-static noise is the only dominant mechanism, which is not generally valid under our experimental conditions. Under continuous driving, additional fast and non-stationary noise channels become increasingly important: phonon-assisted processes, laser-induced charge trapping, and electric-field–induced g-factor fluctuations, which grow in importance with increasing magnetic field. The coexistence of multiple noise sources with different correlation times naturally leads to \(\alpha < 2\).

Based on these considerations and the fitted values of \(\alpha\) from cw-g2 measurements varying the laser power and magnetic field (cf.~\ref{sec:app_cw_g2_power_series},~\ref{sec:app_cw_g2_bfield_series}) we chose to set \(\alpha = 1\) when fitting Eq.~\ref{eq:fit_function_pulsed_docp} to the data presented in Fig.~\ref{fig:pulsed_coherence}. 

\newpage
\subsubsection{Spin dephasing and precession frequency depending on readout-time}

\begin{figure}[hbt!]
    \centering
    \includegraphics[width=.99\textwidth]{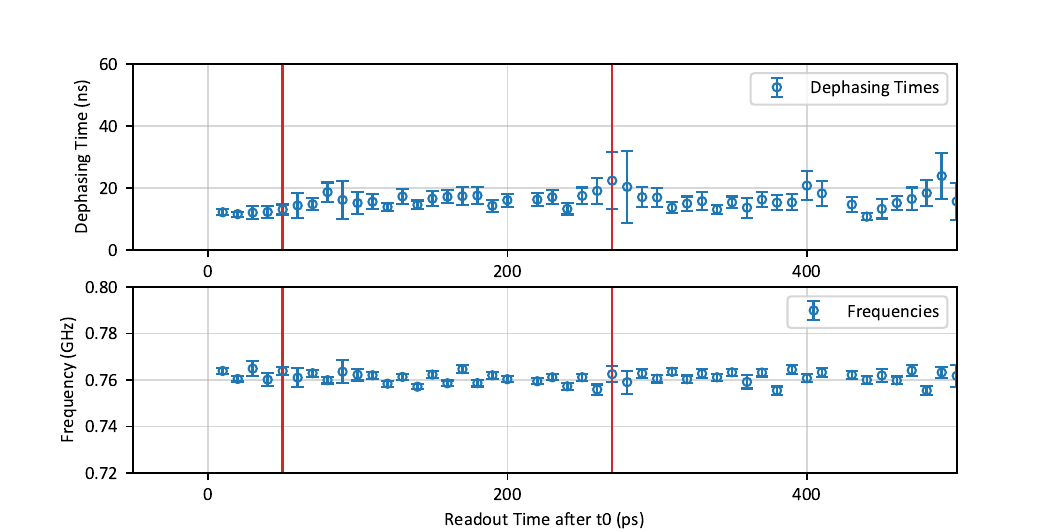}
    \caption{
    \textbf{$\tau$ and $f$ vs. readout time after \(t_{0}\).}
    The red vertical lines mark the start and end point, used for averaging over $\tau$ and $f$. The left value is chosen at \SI{50}{\pico \second} after \(t_{0}\) to start at the point of maximum PL emission. The right value is chosen at \(t_{readout}=270\) in order to average over one full precession period of the excited state (\SI{220}{\pico \second} at \(B_{x}=\)~\SI{150}{\milli \tesla}). Averaging over the dephasing times yields $T_2^*=(15.9\pm1.7)$~\SI{}{\nano \second} and averaging over $f$ yields $f=0.761 \pm 0.002$~\SI{}{\giga \hertz}.
    }
    \label{fig:app_pulsed_decoherence}
\end{figure}

Fig.~\ref{fig:app_pulsed_decoherence} shows a plot of the dephasing times and frequencies of the fitted \(\langle S_z \rangle\) from the pulsed measurement for different readout times.

